\documentclass[twocolumn,letterpaper,aps,prc,longbibliography,nofootinbib,superscriptaddress,showpacs,floatfix]{revtex4-1}
\usepackage[english]{babel}
\usepackage[colorlinks,linkcolor=blue,citecolor=blue,filecolor=black,urlcolor=blue]{hyperref}
\usepackage{amsmath}
\usepackage{graphicx}
\newcommand{\pT} {p_{\mathrm{T}}}

\begin{document}
\title{Impact of nuclear structure on longitudinal flow decorrelations in high-energy isobar collisions}

\author{Maowu Nie}\affiliation{Institute of Frontier and Interdisciplinary Science, Shandong University, Qingdao, 266237, China}\affiliation{Key Laboratory of Particle Physics and Particle Irradiation, Ministry of Education, Shandong University, Qingdao, Shandong, 266237, China}
\author{Chunjian Zhang}\affiliation{Department of Chemistry, Stony Brook University, Stony Brook, New York 11794, USA}
\author{Zhenyu Chen}\affiliation{Institute of Frontier and Interdisciplinary Science, Shandong University, Qingdao, 266237, China}\affiliation{Key Laboratory of Particle Physics and Particle Irradiation, Ministry of Education, Shandong University, Qingdao, Shandong, 266237, China}
\author{Li Yi}\affiliation{Institute of Frontier and Interdisciplinary Science, Shandong University, Qingdao, 266237, China}\affiliation{Key Laboratory of Particle Physics and Particle Irradiation, Ministry of Education, Shandong University, Qingdao, Shandong, 266237, China}
\author{Jiangyong Jia}\affiliation{Department of Chemistry, Stony Brook University, Stony Brook, New York 11794, USA}\affiliation{Physics Department, Brookhaven National Laboratory, Upton, New York 11976, USA}

\begin{abstract}
Fluctuations of harmonic flow along pseudorapidity, known as flow decorrelations, are an important probe of the initial state geometry of the quark-gluon plasma. The flow decorrelations are shown to be sensitive to the collective structure of the colliding nuclei, as revealed clearly by comparing collisions of isobars, $^{96}$Ru+$^{96}$Ru and $^{96}$Zr+$^{96}$Zr, which have different nuclear structures. The flow decorrelations in central collisions are mostly sensitive to nuclear deformations, while those in the mid-central collisions are primarily sensitive to differences in skin thickness between  $^{96}$Ru and $^{96}$Zr. Longitudinal flow decorrelations in heavy-ion collisions are a new tool to probe the structure of colliding nuclei.
\end{abstract}
\maketitle

\section{Introduction}
Azimuthal anisotropic flow~\cite{Ollitrault:1992bk} is an important tool for studying the properties of the quark-gluon plasma produced in high-energy heavy-ion collisions at the Relativistic Heavy Ion Collider (RHIC) and the Large Hadron Collider (LHC)~\cite{Adams:2005dq, Adcox:2004mh, Muller:2012zq}. The flow is characterized through a Fourier expansion of particle production in each event, where $dN/d\phi \propto 1+ 2\sum_{n=1}^{\infty} v_{n} \cos n(\phi - \Psi_{n})$, with $v_n$ and $\Psi_n$ representing the magnitude and phase of the $n^{th}$-order of flow, respectively. The dominant components, $v_2$ (elliptic flow)~\cite{Voloshin:1994mz} and $v_3$ (triangular flow)~\cite{Alver:2010gr}, have been extensively studied, providing constraints on the initial condition and the transport properties of the QGP~\cite{Bernhard:2016tnd,Nijs:2022rme,Everett:2020xug}.

One important insight that has recently emerged is that the initial condition for heavy-ion collisions is not boost-invariant in the longitudinal direction. Instead, it interpolates between the geometry of the projectile nucleus in the forward direction and the geometry of the target nucleus in the backward direction. These two geometries differ due to random fluctuations of participating nucleons, resulting in a twist in the final state event-plane angles or an asymmetry in the flow magnitudes~\cite{Bozek:2010vz, Bzdak:2012tp, Xiao:2012uw, Huo:2013qma, Jia:2014ysa}. Such flow decorrelations have been measured at both the LHC~\cite{CMS:2015xmx,ATLAS:2017rij,ATLAS:2020sgl} and RHIC~\cite{Nie:2019bgd,Nie:2020trj}, and they are described by 3+1D event-by-event viscous hydrodynamical model calculations~\cite{Denicol:2015nhu,Bozek:2015bna,Bozek:2015bha,Pang:2014pxa, Pang:2015zrq,Wu:2018cpc,Pang:2018zzo}. 

The experimental observable for flow decorrelations is constructed as a ratio of two-particle correlations
\begin{align}\label{eq:1}
r_n(\eta, \eta_{\mathrm{ref}}) = \frac{V_{n\Delta}(-\eta,\eta_{\mathrm{ref}})}{V_{n\Delta}(\eta,\eta_{\mathrm{ref}})}\;.
\end{align}
Here, $\eta_{\mathrm{ref}}$ is the reference pseudorapidity common to the numerator and the denominator, typically chosen at large rapidity. The decorrelations are reflected by the factorization breaking, where $V_{n\Delta}(\eta,\eta_{\mathrm{ref}}) \neq v_n(\eta)v_n(\eta_{\mathrm{ref}})$, leading to a linear decrease of $r_n$ as a function of $\eta$. 

The longitudinal fluctuations of the initial condition are also expected to depend on the collective structure of the colliding nuclei~\cite{Jia:2021tzt, Giacalone:2021udy}. The shape and radial profiles of heavy nuclei are often described by a Woods-Saxon form: 
\begin{align}\nonumber
\rho(r, \theta, \phi) &\propto \frac{1}{1+e^{(r-R(\theta, \phi))/a_0}}\\\label{eq:2}
R(\theta, \phi) &= R_0(1+\beta_2Y^0_2( \theta, \phi)+\beta_3Y^0_3( \theta, \phi))\;.
\end{align}
In this form, the nuclear surface $R(\theta, \phi)$ incorporates quadrupole deformation $\beta_2$ and octupole deformation $\beta_3$. $R_0$ and $a_0$ represent the half-height radius and nuclear skin, respectively. Recent studies have shown that the $v_n$ values are enhanced by $\beta_n$, especially in ultra-central collisions (UCC). In mid-central collisions, the $v_n$ values are also influenced by $R_0$ and $a_0$~\cite{Jia:2021oyt}.  The evidence for these influences is best revealed by ratios of observables between two isobaric collision systems, $^{96}$Ru+$^{96}$Ru and $^{96}$Zr+$^{96}$Zr collisions, which have different structures. The ratios of many observables, including $v_2$ and $v_3$~\cite{STAR:2021mii}, exhibit significant departure from unity. Thus, it is interesting to ask the following question: {\it How do the structural differences between $^{96}$Ru and $^{96}$Zr impact the longitudinal structure of the initial condition and subsequently the flow decorrelations?}

This paper examines the influence of nuclear deformations and nuclear radial distribution on flow decorrelations in $^{96}$Zr+$^{96}$Zr and $^{96}$Ru+$^{96}$Ru collisions within a Multi-phase transport (AMPT) model. We found that these structural parameters do impact the flow decorrelations captured by the $r_n$ observables defined in Eq.~\ref{eq:2}. 

\section{Model setup}
We use the AMPT model in the string melting mode~\cite{Lin:2004en} to simulate isobaric $^{96}$Zr+$^{96}$Zr and $^{96}$Ru+$^{96}$Ru collisions at $\sqrt{s_{\mathrm{NN}}}=200$ GeV. In the model, the interaction of initial partons generates strings with varying lengths in the longitudinal direction, resulting in fluctuations in the initial geometry that can reproduce the measured flow decorrelation in heavy-ion collisions~\cite{Ma:2016fve, He:2020xps}. The string melting version of AMPT comprises four main components: Monte Carlo Glauber model for initial conditions, strings, and mini-jet that melt into partons from the HIJING model~\cite{Wang:1991hta}, an elastic parton cascade described by the ZPC model~\cite{Zhang:1997ej}, a quark coalescence model for hadronization, and hadron rescatterings described by the ART model~\cite{Li:1995pra}. The elastic parton-parton cross section is chosen as the standard value of 3 mb at RHIC top energy.

The observable $r_n$ for flow decorrelations in Eq.~\ref{eq:1} is constructed using hadrons with transverse momentum $0.4 < \pT <$ 4 GeV/$c$ as a function of pseudorapidity within the range $|\eta|<1.5$. In experimental measurements, a rapidity gap is often required between $\eta$ and $\eta_{\mathrm {ref}}$ to suppress non-flow correlations associated with jet fragmentation and resonance decays. For this analysis, we choose the reference pseudorapidity to be $3.1 < |\eta_{\mathrm {ref}}| < 5.1$ for $r_2$ and $2.1 < |\eta_{\mathrm {ref}}| < 5.1$ for $r_3$. These gap requirements effectively remove non-flow correlations arising from near-side jet fragmentation, while allowing some non-flow correlations associated with away-side jets to remain partially. However, we found that in the low $\pT$ region considered in this analysis, the away-side jets primarily contribute to $r_2$, while their impact on $r_3$ is negligible. Additionally, we observed that enlarging the rapidity gap beyond the default values does not significantly alter the $r_n$ values.

The impact of nuclear structure for any given observable $\mathcal{O}$, including $r_n$, is revealed by the isobar ratio $R_{\mathcal{O}}$ between Ru+Ru and Zr+Zr collisions:
\begin{align}
R_{\mathcal{O}}& = \frac{\mathcal{O}_{\mathrm{Ru+Ru}}}{\mathcal{O}_{\mathrm{Zr+Zr}}}\;.
\end{align}
A recent study has shown that these isobar ratios are insensitive to final state effects, such as the shear viscosity, hadronization, and hadronic cascade~\cite{Zhang:2022fou}, and primarily reflect the properties of the initial state geometry. 

The Glauber parameters for $^{96}$Ru and $^{96}$Zr in the AMPT model are taken from Ref.~\cite{Zhang:2021kxj}, and are listed in Table~\ref{tab:1}. These parameter sets have been shown to successfully describe the centrality dependence of the isobar ratios for $v_2$, $v_3$, and multiplicity distribution. Specifically, the larger quadupole deformation in Ru, $\beta_{2\mathrm{Ru}} = 0.162> \beta_{2\mathrm{Zr}} = 0.06$, have been shown to increase the ratio $R_{v_2}$ in UCC. The larger octupole deformation in Zr, $\beta_{3\mathrm{Ru}} =0 < \beta_{3\mathrm{Zr}} = 0.2$, have been shown to decrease $R_{v_3}$ in UCC and increase $R_{v_2}$ in mid-central collisions. The smaller nuclear skin $a_0$ in Ru, $a_{0\mathrm{Ru}} =0.46$~fm $< a_{0\mathrm{Zr}} = 0.52$ fm, have been shown to increase $R_{v_2}$ in mid-central collisions, while the small difference in their half-radius $R_0$ have been shown not to cause a difference in the $v_n$ values in the two systems. Following this logic, we aim to study how the differences in each of the four parameters impact the isobar ratios $R_{r_2}$ and $R_{r_3}$. To do this, we introduce three additional sets of Wood-Saxon parameters that turn on the difference in one parameter at a time. Thus, we have five sets of structural parameters for isobar collisions as listed in Table~\ref{tab:1}.

The study of flow decorrelations requires a significant amount of event statistics. We have simulated a large number of events, approximately 100 million events for each of the five parameter sets in Table~\ref{tab:1}. However, the statistical precision is not sufficient for a detailed investigation of the centrality dependence. In this paper, we calculate $r_n$ and $R_{r_n}$ in two broad centrality intervals 0--10\% and 10--40\%.

\begin{table}[htbp]
	\caption{Nuclear structure parameter sets used for $^{96}$Ru (Case1) and $^{96}$Zr (Case5), as well as three intermediate parameter sets (Case2--Case4), which allow us to study the influence of one structure parameter at a time.}
	\begin{ruledtabular} \begin{tabular}{lcccc}
			Model setup & $R_0$ & $a_0$  &  $\beta_2$ & $\beta_3$\\
			\hline
			Case1 $^{96}\mathrm {Ru}$  & 5.09 & 0.46 &0.162 & 0\\
			Case2   & 5.09 & 0.46 &0.06 & 0\\
			Case3   & 5.09 & 0.46 &0.06 & 0.2\\
			Case4   & 5.09 & 0.52 &0.06 & 0.2\\
			Case5 $^{96}\mathrm Zr$   & 5.02 & 0.52 &0.06 & 0.2\\ 
	\end{tabular} \end{ruledtabular}
	\label{tab:1}
\end{table}

\section{Results and discussions}
\begin{figure}[h!]
\centering
\includegraphics[width=1.\linewidth]{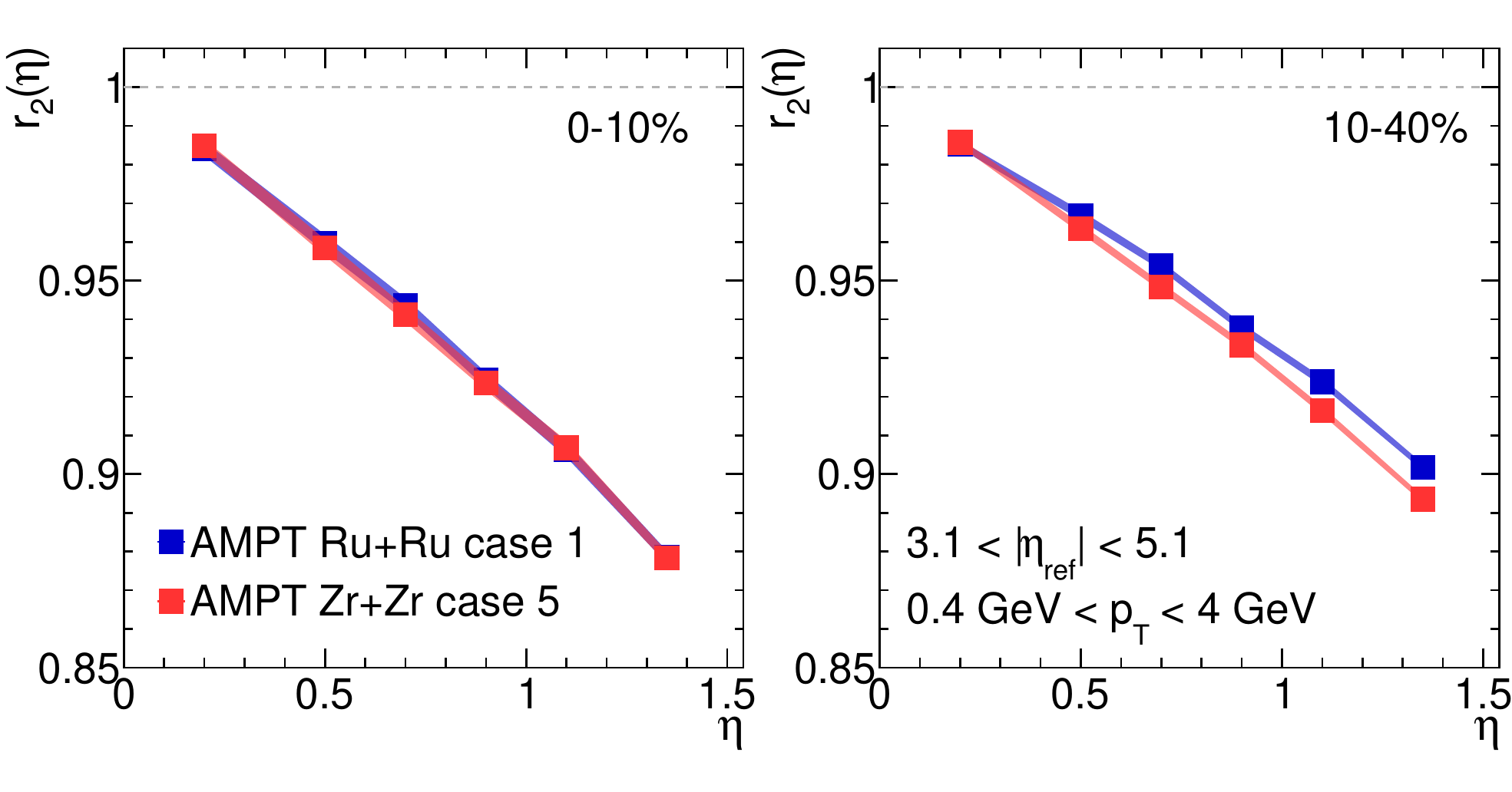}
\includegraphics[width=1.\linewidth]{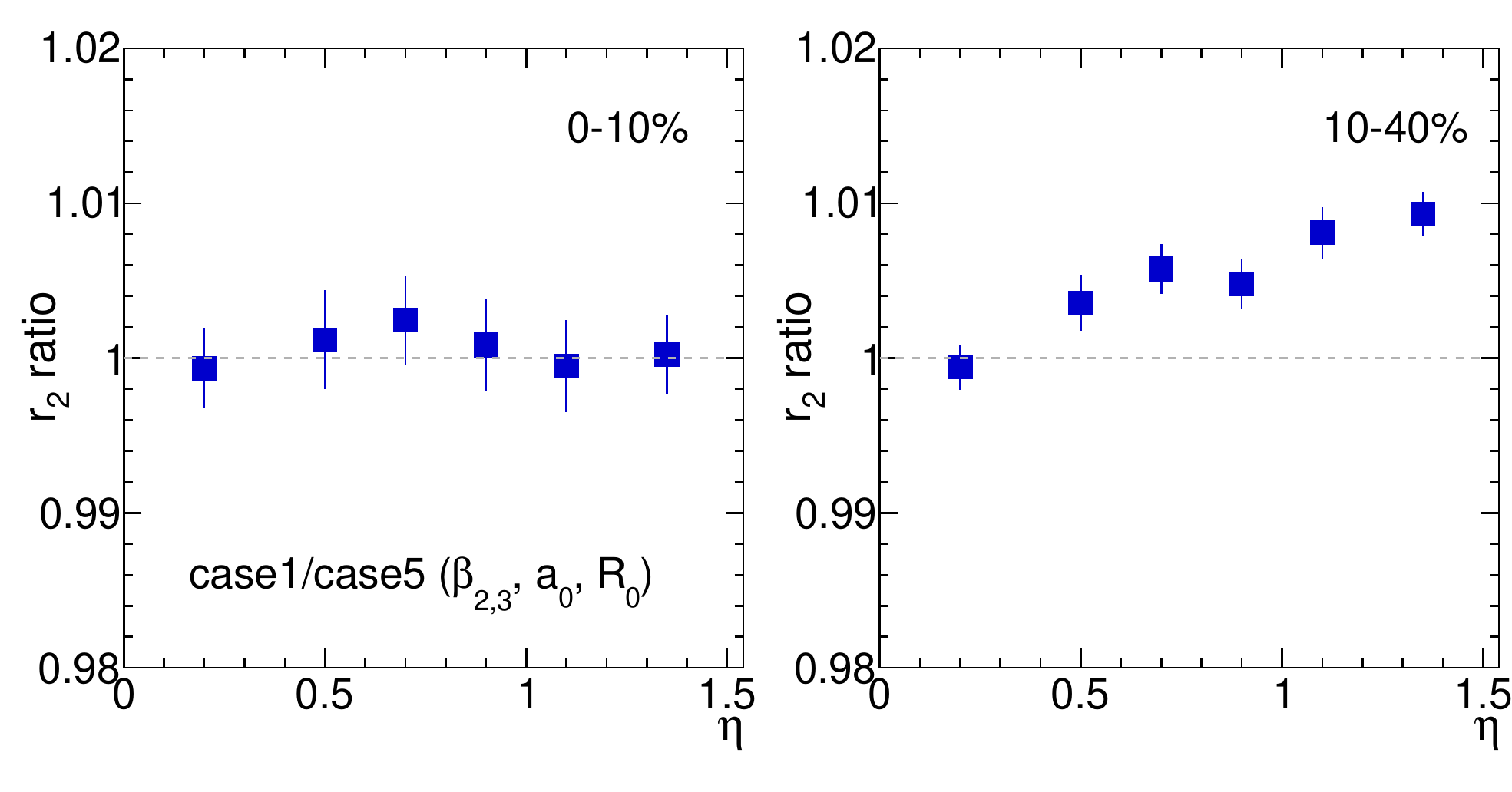}
\caption{$r_2(\eta)$ (top row) in Zr+Zr and Ru+Ru collisions and the ratios $R_{r_2}$ (bottom row) in 0--10\% (left column) and 10--40\% (right column) centrality intervals.}\label{fig:1}
\end{figure}

\begin{figure}[h!]
\centering
\includegraphics[width=1.\linewidth]{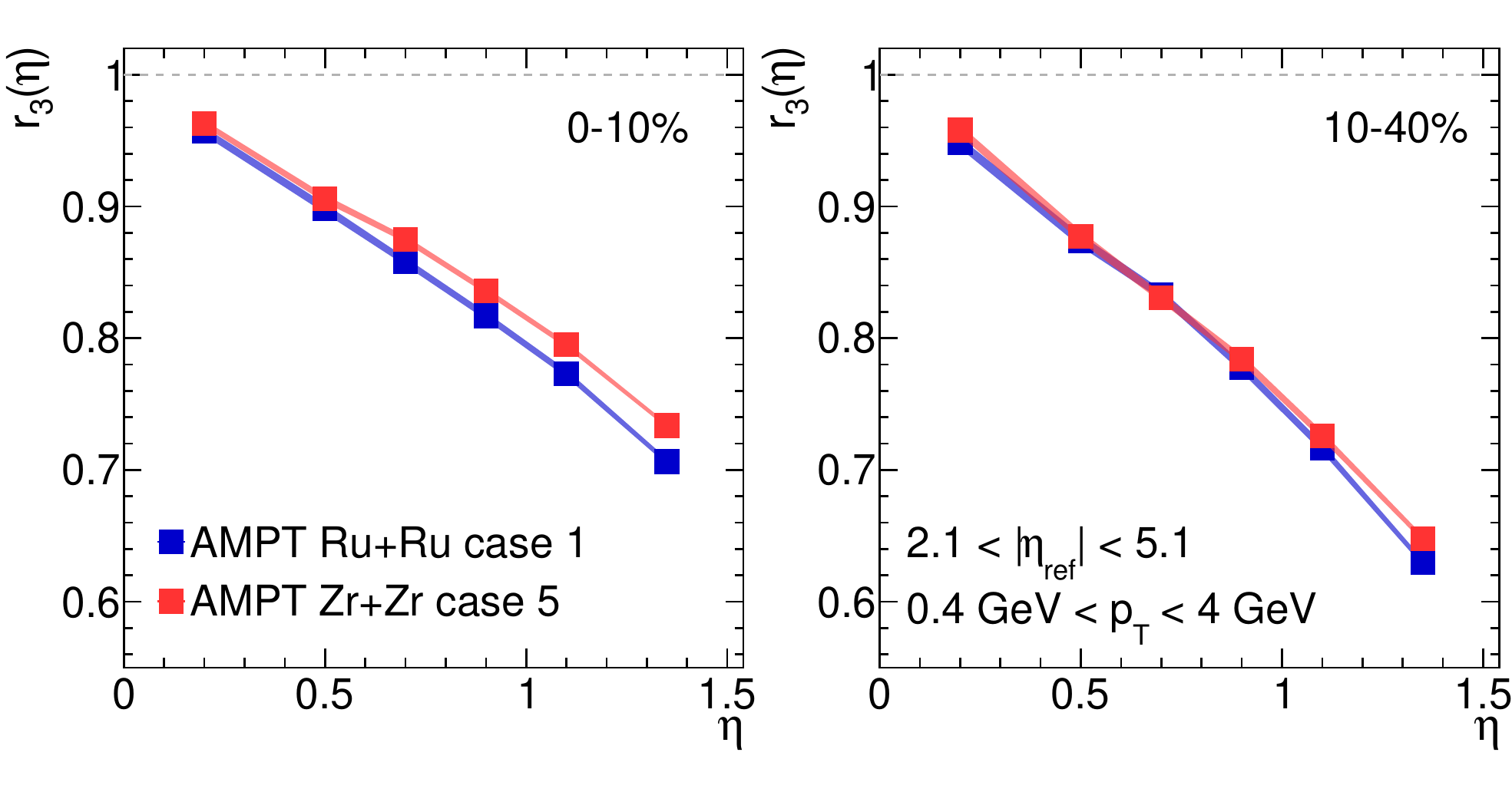}
\includegraphics[width=1.\linewidth]{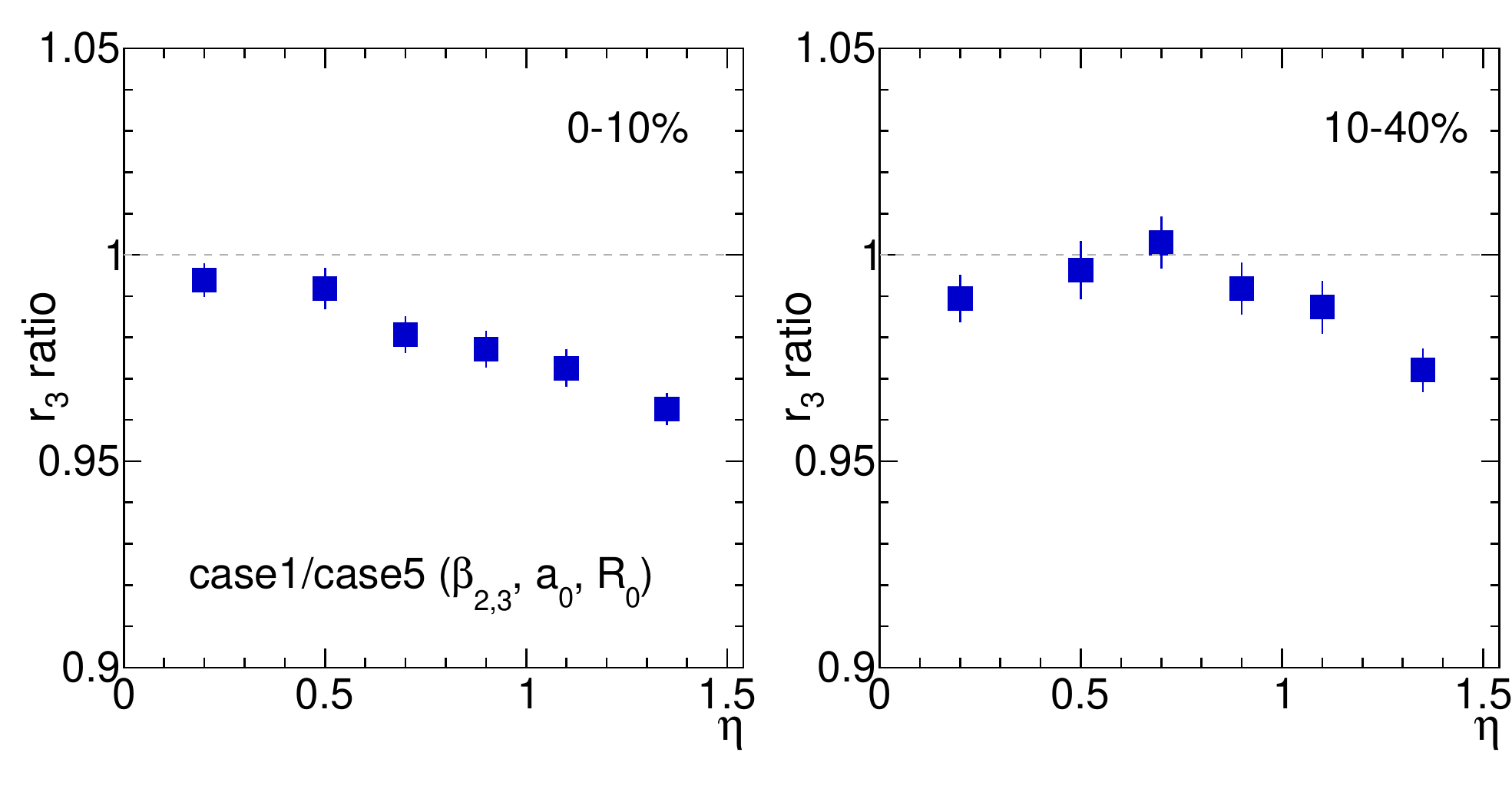}
\caption{$r_3(\eta)$ (top row) in Zr+Zr and Ru+Ru collisions and the ratios $R_{r_3}$ (bottom row) in 0--10\% (left column) and 10--40\% (right column) centrality intervals.}
\label{fig:2}
\end{figure}

Figure~\ref{fig:1} depicts the $r_2$ values and isobar ratio $R_{r_2}$ as a function of $\eta$ in 0--10\% and 10--40\% centrality intervals. The presence of elliptic flow decorrelations reduces the $r_2$ values by approximately 10\% around $\eta= 1.5$, and a clear difference between Ru+Ru and Zr+Zr is observed primarily in the 10--40\% centrality range, where the $r_2$ values in Ru+Ru is approximately 1\% larger. The results for $r_3$ are shown in Fig.~\ref{fig:2}. The overall level of decorrelations is much larger, reaching about 30\% around $\eta=1.5$. The difference between the isobars is clearly observed in the 0--10\% centrality interval, where the $r_3$ values in Ru+Ru are approximately 4\% smaller.

To examine how these differences are impacted by each of the four Wood-Saxon parameters individually, we performed a step-by-step variation between different cases, with each step considering only the impact of one parameter at a time. For instance, the ratio Case2/Case1,  Case3/Case2,  Case4/Case3, or Case5/Case4 includes only the difference in $\beta_2$, $\beta_3$, $a_0$, or $R_0$, respectively. 
\begin{figure}[h!]
  \centering
\includegraphics[width=1.\linewidth]{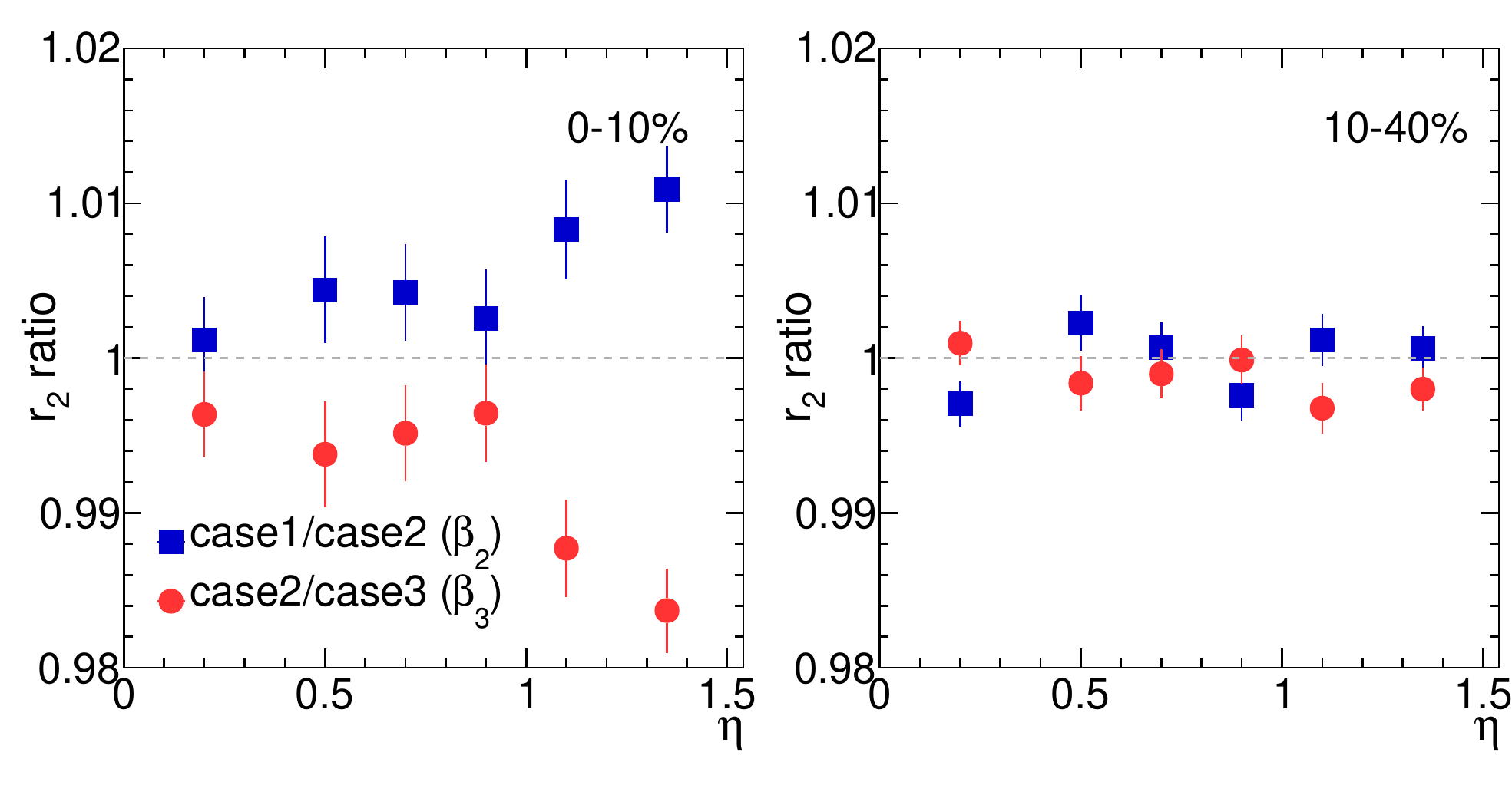}
\includegraphics[width=1.\linewidth]{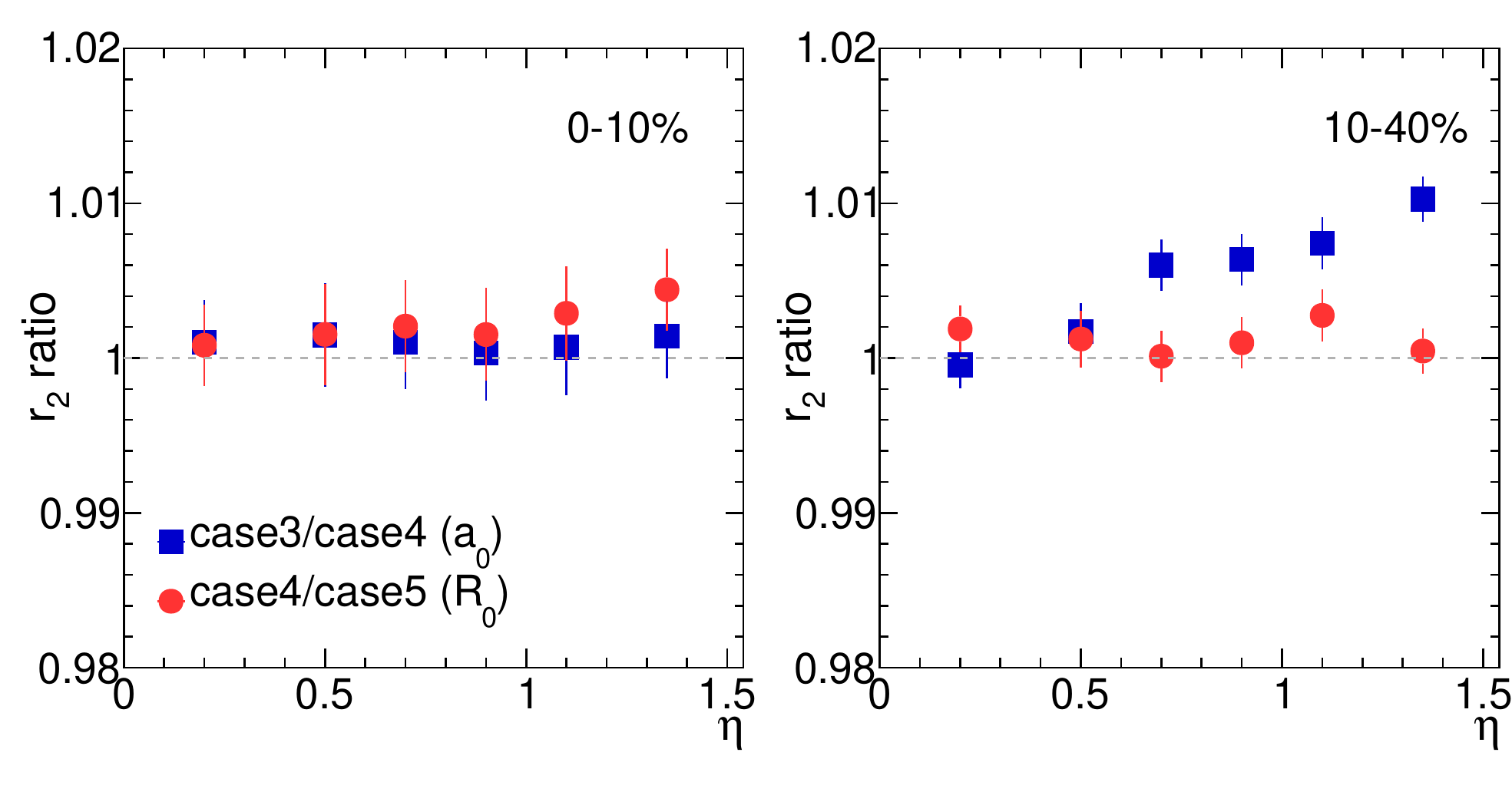}
  \caption{Ratio of $r_2$ including effects of nuclear deformations (top row) and radial profile (bottom row) in 0--10\% (left column) and 10--40\% (right column) centrality intervals.}
  \label{fig:3}
\end{figure}

The ratios of $r_2$ are displayed in Fig.~\ref{fig:3}. We observe that $\beta_2$ and $\beta_3$ only impact $r_2$ in central collisions (top panels), but their contributions almost cancel each other. Specifically, $\beta_2$ enhances the ratio, corresponding to smaller decorrelation in Ru+Ru collisions, while $\beta_3$ decreases the ratio, corresponding to larger decorrelation in Ru+Ru collisions. To understand this, it is sufficient to note that nuclear shape leads to a global variation in collision geometry. Therefore, if the global $v_2$ is enhanced by nuclear deformation, the relative impact on decorrelations naturally becomes smaller. Previous studies have shown that $\beta_2$ ($\beta_3$) enhances (decreases) the $v_2$~\cite{Zhang:2021kxj,Jia:2021oyt}, which is expected to reduce (enhance) the decorrelation of $v_2$.

The impact of $a_0$ and $R_0$ in isobar collisions is shown in the bottom panels of Fig.~\ref{fig:3}. The only noticeable impact is seen for $a_0$ in the 10--40\% centrality interval, where the smaller $a_0$ in Ru is found to increase $v_2$, thus leading to a smaller decorrelation reflected by a larger $r_2$ ratio. Therefore, we conclude that the similarity of isobar ratio $R_{r_2}$ in the 0--10\% centrality interval in Fig.~\ref{fig:1} arises from a fortuitous cancellation of $\beta_2$ and $\beta_3$, while the enhancement of $R_{r_2}$ in the 10--40\% centrality interval is predominantly influenced by the $a_0$ difference between Ru and Zr.  

\begin{figure}[h!]
  \centering
\includegraphics[width=1.\linewidth]{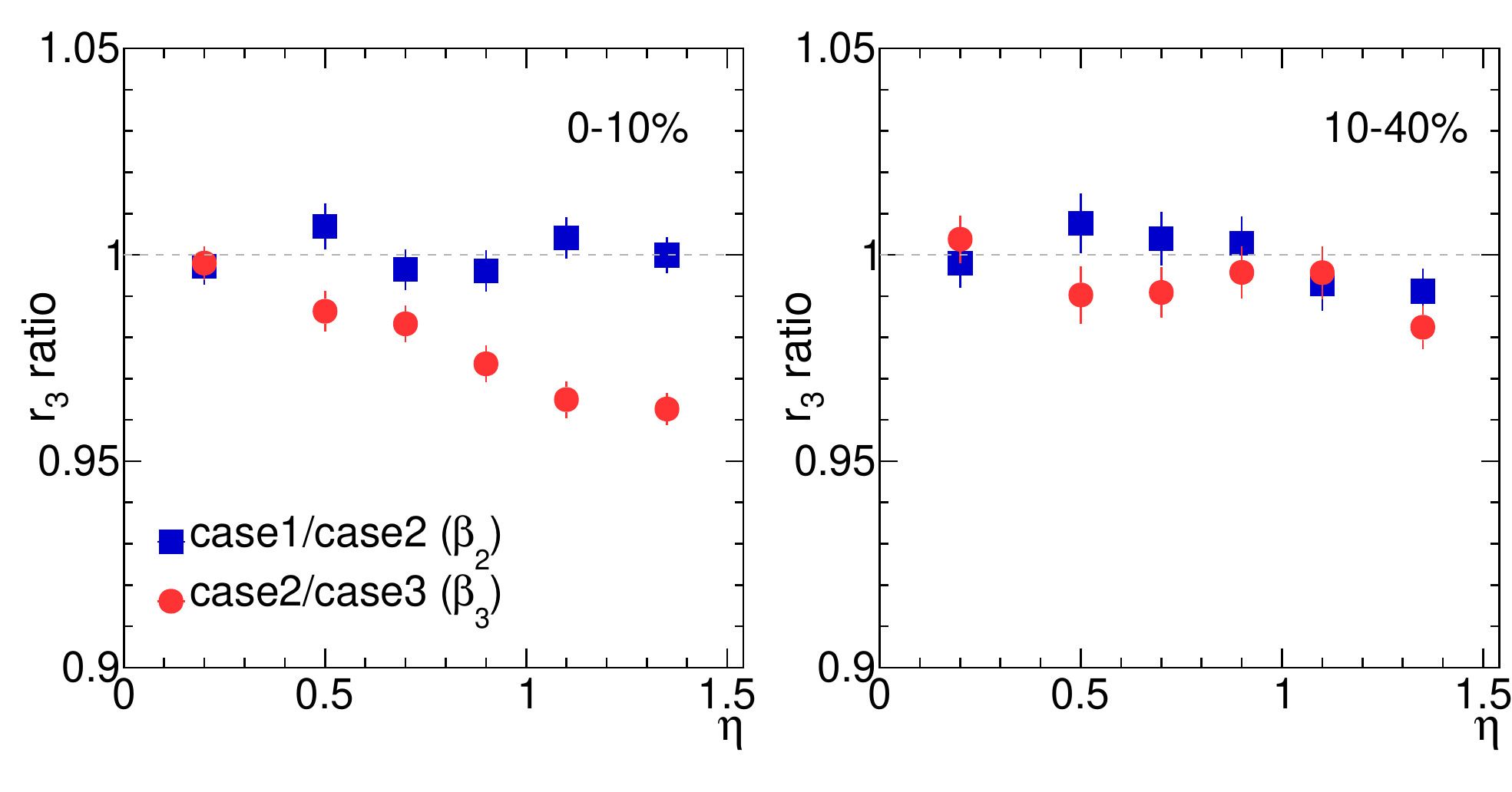}
\includegraphics[width=1.\linewidth]{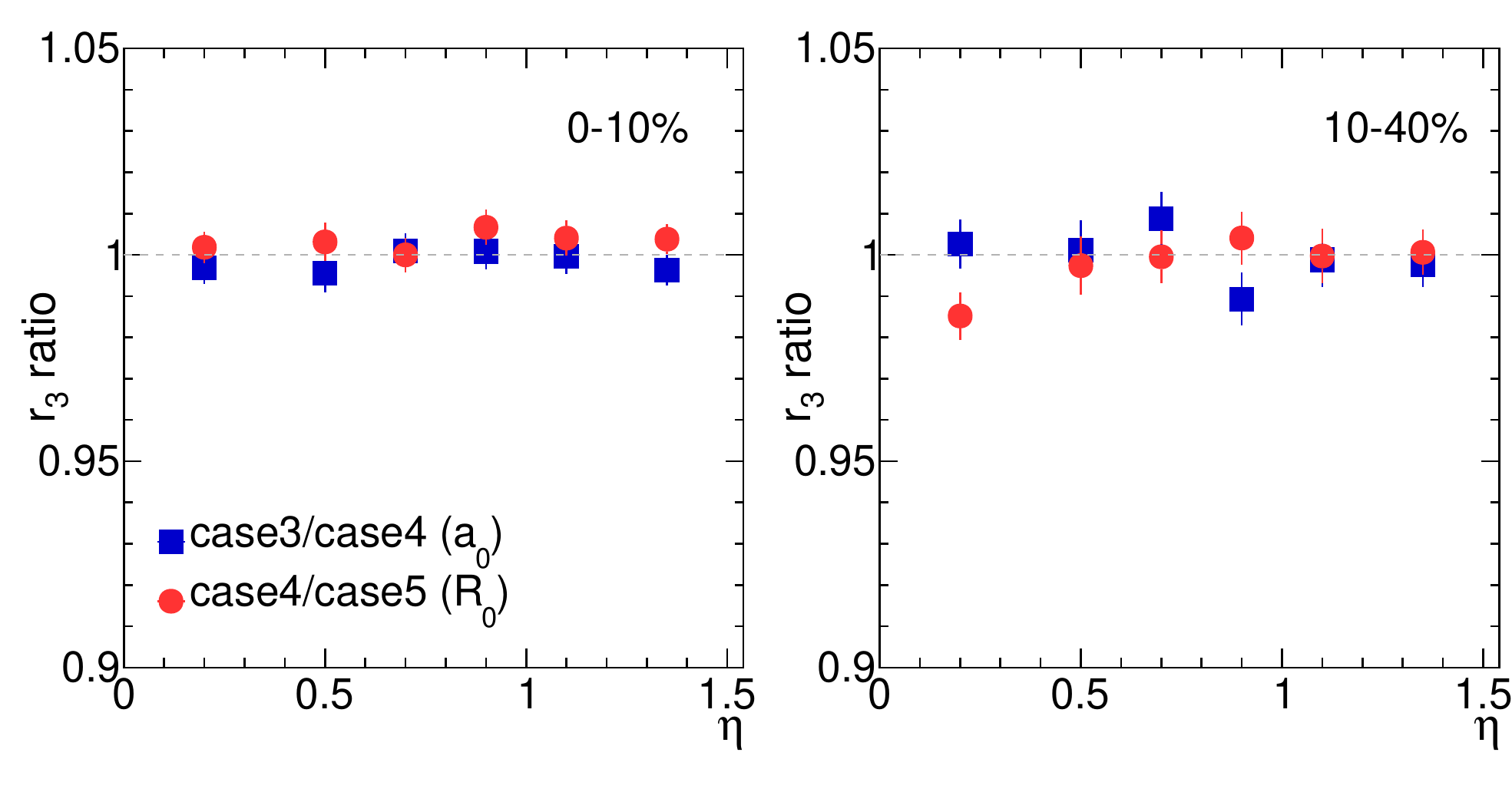}
  \caption{Ratio of $r_3$ including effects of nuclear deformations (top row) and radial profile (bottom row) in 0--10\% (left column) and 10--40\% (right column) centrality intervals.}
  \label{fig:4}
\end{figure}

The impact of nuclear structure on isobar ratios for $r_3$ is depicted in Fig.~\ref{fig:4}. It is evident that the deviation of $R_{r_3}$ from unity is dominated by $\beta_3$ in Zr, which reduces the ratio. This impact is most pronounced in the 0--10\% most central collisions but is also noticeable at a lower level in the 10--40\% centrality interval. The interpretation of this is similar to that for $r_2$: the larger $\beta_3$ globally increases the $v_{3\mathrm{Zr}}$, thereby reducing its decorrelation and resulting in a value of $R_{r_3}$ smaller than one.

\section{Conclusion}
The impact of nuclear structure on longitudinal flow decorrelations in Ru+Ru and Zr+Zr collisions is investigated using a transport model. The quadrupole deformation $\beta_2$ is found to decrease the decorrelation for elliptic flow $v_2$, while the octupole deformation $\beta_3$ decreases the decorrelation for both elliptic and triangular flow $v_3$. However, the influence of $\beta_n$ is significant only in central collisions. Specifically, the large $\beta_{2\mathrm{Ru}}$ decreases the decorrelation of $v_2$ in Ru+Ru collisions, whereas the large  $\beta_{3\mathrm{Zr}}$ reduces the decorrelation for both $v_2$ and $v_3$ in Zr+Zr collisions. This leads to a fortuitous similarity in the decorrelation of $v_2$ between the two collision systems and a weaker decorrelation for $v_3$ in Zr+Zr collisions. The nuclear skin parameter $a_0$ only impacts the decorrelation of $v_2$ in mid-central collisions. The smaller $a_{0\mathrm{Ru}}$ value results in a smaller decorrelation of $v_2$ in Ru+Ru compared to Zr+Zr collisions. These findings demonstrate that longitudinal dynamics are sensitive to the nuclear structure, providing a valuable tool for extracting nuclear structure using heavy-ion collisions.

{\bf Acknowledgements:} We thank Jianing Dong for maintaining the high-quality performance of the computer facility. M. Nie, L.Yi and Z. Chen are supported by the National Natural Science Foundation of China under Grant No. 12105156, No. 11890710 and No. 11890713, National Key R\&D Program of China under Grant No. 2022YFA1604903 and Shandong Provincial Natural Science Foundation under Grant No. ZR2021QA084. C. Zhang and J. Jia are supported by DOE Award No. DEFG0287ER40331.

\bibliographystyle{apsrev4-1}
\bibliography{sample}

\end{document}